\documentclass[twocolumn,showpacs,aps,epsfig,nofootinbib]{revtex4}
\usepackage{amssymb, amsmath, bm, dcolumn, epsf, graphicx, latexsym, mathbbol, slashed}
\usepackage{graphicx}
\usepackage{dcolumn}
\usepackage{bm}
\usepackage{amssymb}
\usepackage{color}
\usepackage{multirow}
\usepackage[colorlinks,linkcolor=blue,citecolor=blue,urlcolor=blue]{hyperref}

 \def\tskip{\setlength{\tskip}{5pt}}
\def\colwidth{\setlength{\colwidth}{3.5in}}

\newcommand{\lsim}{\mathrel{\hbox{\rlap{\lower.55ex\hbox{$\sim$}} \kern-.3em \raise.4ex \hbox{$<$}}}}
\newcommand{\gsim}{\mathrel{\hbox{\rlap{\lower.55ex\hbox{$\sim$}} \kern-.3em \raise.4ex \hbox{$>$}}}}
\newcommand{\beq}{\begin{equation}}
\newcommand{\eeq}{\end{equation}}
\newcommand{\be}{\begin{equation}}
\newcommand{\ee}{\end{equation}}
\newcommand{\bes}{\begin{equation*}}
\newcommand{\ees}{\end{equation*}}
\newcommand{\beqa}{\begin{eqnarray}}
\newcommand{\eeqa}{\end{eqnarray}}
\newcommand{\bea}{\begin{eqnarray}}
\newcommand{\ena}{\end{eqnarray}}

\usepackage{color}

\begin{document}

\title{Model-independent measurement of the absolute magnitude of Type Ia Supernovae with gravitational-wave sources}

\author{Wen Zhao$^{1,2}$}

\author{Larissa Santos$^{1,2,3}$}
\affiliation{$^{1}$ CAS Key Laboratory for Researches in Galaxies and Cosmology, Department of Astronomy, University of Science and Technology of China, Chinese Academy of Sciences, Hefei, Anhui 230026, China \\
$^{2}$ School of Astronomy and Space Science, University of Science and Technology of China, Hefei 230026, China \\
$^{3}$ Center for Gravitation and Cosmology, College of Physical Science and Technology, Yangzhou University, Yangzhou, 225009, China}



\begin{abstract}
{The similarity of the absolute luminosity profiles of Type Ia supernovae (SNIe), as one kind of distance indicator, has led their use in extragalactic astronomy as secondary standard candles. In general, the empirical relationship of SNIa on the absolute peak magnitude $M_{\rm B}$ is calibrated by Cepheid variables in the near distance scale and directly extrapolated to much
farther distances. Therefore, two main problems arise. First of all, their calibration, in particular the
determination of $M_{\rm B}$, depends on the empirical relationship of Cepheid variables,
which suffers from various uncertainties. The second is related to the
homogeneity of SNIa in their true $M_{\rm B}$, which is known to be
poor in different environments. The observed gravitational-wave
(GW) signal of the coalescence of compact binary systems and their
electromagnetic counterparts provide the novel and
model-independent way to address these two problems. In the era of
second-generation GW detectors, the low-redshift GW sources
provide a novel method to calibrate the
empirical relationship of SNIa, using their self-calibrated
distances. In this paper, we use the event
GW170817 to calibrate the empirical relationship in different
low redshift ranges, and find that the calibration results
are consistent with the ones derived from the Cepheid variables. Moreover,
the uncertainties of $M_{\rm B}$ in both methods are also
comparable. By the observations of third-generation GW detectors, GW sources
can also be used to measure the values of $M_{\rm B}$ for the
high-redshift SNIe, which provides a unique opportunity to study the dependence of $M_{\rm B}$ on the local environment, strength of gravity, and the intrinsic properties of the explosion, in addition to test the homogeneity of standard candles. We find that the uncertainties of $M_{\rm B}$ in both high and low redshifts are more than one order of magnitude smaller than the current accuracy.




}
\end{abstract}


\maketitle

\section{Introduction \label{section1}}


Cosmic distance ladder plays a key role in determining the
distances to celestial objects, which is a succession method
depending on the empirical relations for various specific stars or
galaxies. In general, these empirical relations are calibrated in
the low redshift (i.e. low-$z$) range, and directly extrapolated
to higher-$z$ ranges. The stabilities of these extrapolations
are always difficult to be directly tested, which may induce to
systematical biases in determining the distance in high-$z$, misleading the explanation of the cosmological data. For instance, based on the similarity of the absolute luminosity profiles, type Ia
supernovae (SNIe) are the best way to determine extragalactic
distances \cite{SNIa-review}. Even though the value of their peak magnitude is
calibrated by the nearby Cepheid variables, the application is assumed to be in a
much farther distance range \cite{riess2009}. Both steps suffer from various
uncertainties. First, the calibration depends on the empirical
period-luminosity relation of Cepheid variables \cite{phillipse}. However, the impact of metallicity on both the zero-point and the slope of this relation, the effects of photometric contamination and a
changing extinction law on Cepheid distances are actively debated
in literature \cite{ceipheid-review}. Meanwhile, the homogeneity of SNIa in their true
$M_{\rm B}$ is proved to be poor in different environments \cite{poor}. The formation of SNIe is still in debate, but in the double degenerate scenario, SNIe are triggered by the merger of two double dwarfs whose mass exceeds the Chandrasekhar limit, raising questions about the applicability of SNIe as standard candles.  The total mass of the two merging white dwarfs varies significantly, meaning that the luminosity also varies \cite{double}. In addition, recent studies show
that in the modified gravity scenario, proposed as a way to interpret the
cosmic acceleration, the peak luminosity of SNIa depends on the
local strength of gravity \cite{LiBaojiu2017,zhao2018}. Therefore, the extrapolation of SNIa's
absolute magnitude from low-$z$ to high-$z$ is dangerous, being the measurement of the absolute magnitude of SNIe in different redshifts necessary. This is of crucial importance in order to study the intrinsic physics, the environment dependence and the local gravitational properties of SNIe.


The observation of gravitational-waves (GW) signal, caused by inspiralling
and merging binary neutron stars (BNSs) as well as neutron star-black
hole binaries, provides a novel to measure the luminosity distance
in an absolute way, without having to rely on a cosmic distance
ladder \cite{schutz}. In many
cases, it is also possible to identify their electromagnetic (EM) counterparts and determine their redshifts \cite{shGRB}.
Therefore, GW standard sirens provide a unique model-independent way to construct the Hubble diagram in a wide redshift
range. Considering the second-generation (2G) GW detector network,
only the BNSs at $z<0.1$ are expected to be observed, which provide an alternative model-independent way to calibrate the SNIa standard candle. The third-generation
(3G) detector network can detect the GW signals of BNSs if
$z\lesssim 2$ \cite{et,ce}, which provides a powerful way to calibrate
various traditional distance indicators, including SNIe. More
importantly, the GW Hubble diagram provides a unique opportunity to
directly test the homogeneity of cosmic distance ladder, and
investigate the intrinsic properties of the celestial objects. For
instance, the direct measurement of the absolute magnitude of SNIa in
different redshift ranges can infer their local strength of
gravity,  becoming a powerful probe of gravity \cite{LiBaojiu2017,zhao2018}.


In this article, we first use the observed event GW170817 to
calibrate the absolute magnitude and the empirical relation of SNIe. We find that the GW calibration, in comparison with the traditional Cepheid variables one, gives consistent results. In addition, we find that the accuracies of both calibration methods are comparable, even if only
one GW source is considered. In the forecast, we consider the
3G detector network, consisting of the Einstein Telescope (ET) and the Cosmic
Explorer (CE), to investigate the potential measurement accuracies of
the SNIa's absolute magnitude in both low-$z$ and high-$z$ ranges \footnote{Similar idea is also investigated in the recent work \cite{sathya2019}.}.
Moreover, we also study the possibility of using GW sources as
cosmology-independent calibration of empirical luminosity relation of $\gamma$-ray bursts (GRBs) \footnote{Similar idea is also investigated in the recent work \cite{wangfy}.}.

\section{Calibrating the empirical relation of SNIa with GW170817}


In the Friedmann-Robertson-Walker universe, the Hubble constant is $H_0=100h_0~{\rm km~s}^{-1}{\rm~Mpc}^{-1}$, where $h_0=0.674\pm 0.005$ was recently derived from the cosmic
microwave background radiation (CMB) data \cite{planck2018}.
Meanwhile, the SNIa data inferred a different value of $h_0=0.7403\pm 0.0142$ \citep{riess2019}, which conflicts with the CMB result at more than 4$\sigma$ confidence
level. The tension in different measurements of the $h_0$ value puzzles modern
cosmology. Many authors argued that the Cepheid calibration for
SNIa might be responsible for this discrepancy. In order to solve these debates, and test the reliability of SNIa as
cosmic distance indicator, it is urgent to search for an independent
calibration method. On the other hand, observing GWs from binary
coalescences at low-$z$ can help determine the sources'
luminosity distance at reasonably high precisions. The discovery
of GW event GW170817, which is caused by the merge of binary
neutron stars, as well as the identification of its various EM
counterparts, open the new era of GW astronomy \cite{GW170817PRL}. In addition to
directly study the cosmic evolution by GW data alone, the GW
standard sirens also provide a novel and model-independent way
to calibrate (or test) the existing cosmic distance ladder,
including SNIa and GRB. In this section, we use the observed GW170817 to calibrate the empirical relation of low-$z$ SNIa, in particular to determine their absolute magnitude. For this GW event, we have the information of its redshift and luminosity distance, which are $z=0.0103$, and
$d=43.8_{-6.9}^{+2.9}{\rm Mpc}$ \cite{GW170817Nature}. For SNIe data, we adopt the observations obtained by SDSS-II and SNLS collaborations \cite{SNIa-data}, which include 740 SNIa data at $z_{\max}<1.3$. The empirical relation of SNIa can be formally written as \cite{SNIa-data}
$S=a+bX_1+c\mathcal{C}+Z$,
where $S\equiv 5\log_{10}(d/{\rm 10pc})$, $a\equiv -M_{\rm B}$ with $M_{\rm B}$ the corrected absolute magnitude of SNIa. The nuisance parameters are described by $a$, $b$ and $c$. $X_1$ stands for the time stretching of the light-curve, and $\mathcal{C}$  describes the supernova color at maximum brightness.
$Z \equiv m_{\rm B}^*-\Delta M_B$, where $m^*_{\rm B}$ corresponds to the apparent magnitude at time of $B$-band maximum. $\Delta M_{\rm B}=-0.08$ if the host galaxy mass of SNIa is larger than $10^{10}{\rm M}_{\odot}$, and zero otherwise.

In the low-$z$ range, i.e. $z\ll1$, up to the second order of redshift, the extended Hubble law is satisfied for any cosmological model, which reads \cite{weinberg}
$d(z)=({z}/{H_0})[1+{1}/{2}(1-q_0)z]$,
where $d(z)$ is the luminosity distance at redshift $z$, $q_0$ is the deceleration parameter. The basic idea is to determine the $H_0$ and $q_0$ by GW data, and construct the Hubble diagram. Then, we calibrate the SNIa empirical relation by this Hubble diagram. This is equivalent to simultaneously fit the five parameters together ($a,b,c,h_0,q_0$) by combining GW and SNIa data, which can be carried out by the usual $\chi^2$ analysis.

We first illustrate if it is possible to auto-calibrate the empirical relation by SNIa data alone. Using the SNIa data at $z<0.05$, and employing the revised CosmoMC package, we obtain the marginalized constraints on the five parameters, which are listed in Table \ref{table1}. We find that the uncertainties of the absolute luminosity, the Hubble constant and the deceleration parameter are all large, which indicates that the auto-calibration of the empirical relation is impossible. The traditional calibration method uses Cepheid observations, which share the same host galaxies with some SNIa in low-$z$ range. However, as mentioned above, this Cepheid calibration depends on other calibration methods in the lower redshift, i.e. the trigonometric parallax method. In addition, the Cepheid calibration also suffers from various uncertainties.

\begin{figure}
\begin{center}
\centerline{\includegraphics[width=8cm,height=4.5cm]{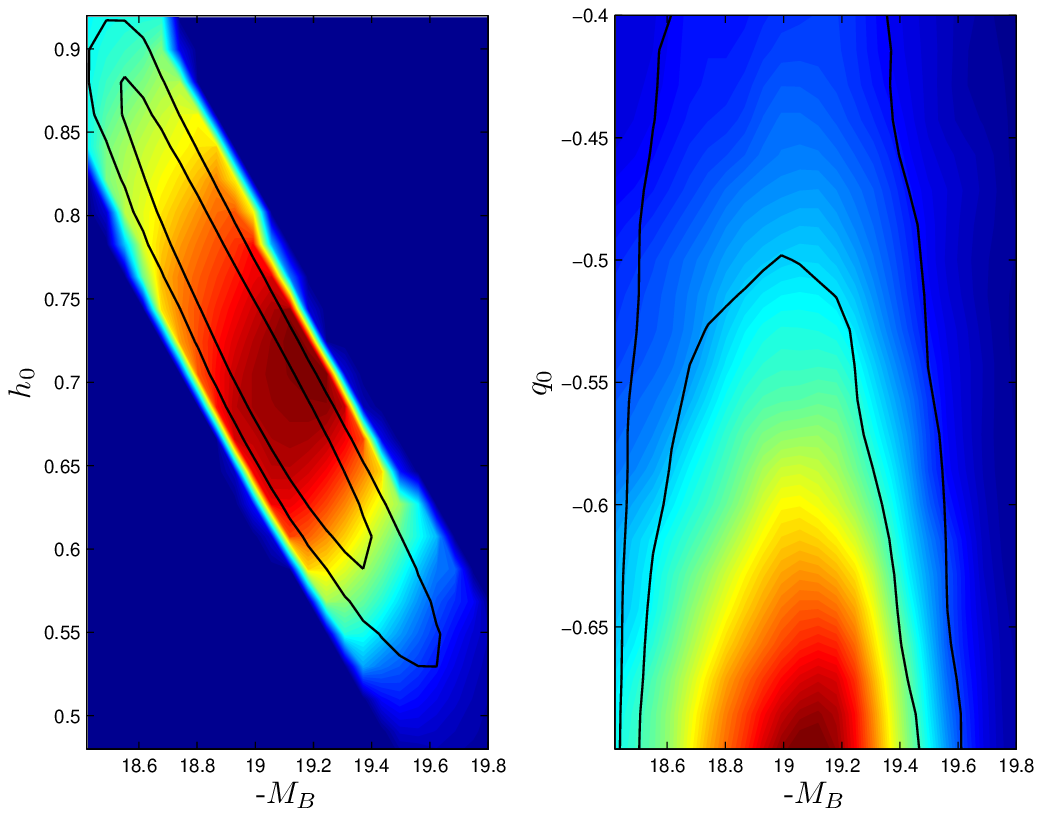}}
\end{center}\caption{Two dimensional marginalized constraints in $M_{\rm B}$-$h_0$ (left) and $M_{\rm B}$-$q_0$ (right) from the data combination of GW170817 and SNIe at $z<0.1$. }\label{f1}
\end{figure}

In this article, we calibrate the empirical relation by combining the GW170817 and the low-$z$ SNIa data. The total $\chi^2$ is the sum of $\chi^2_{\rm GW}$ and $\chi^2_{\rm SN}$. Adopting the SNIa in different low-$z$ ranges, from $z_{\max}=0.05$ to $z_{\max}=0.2$, we derive the constraints on the parameter set, which are all listed in Table \ref{table1}. We find that the constraints of all the parameters are nearly the same for the case with SNIa at $z<0.05$ and those at $z<0.1$. The uncertainty of $M_{\rm B}$ reduces to around $0.3$ magnitude, becoming comparable with the traditional Cepheid calibration method, which follows the calibration accuracy $\sigma_{M_{\rm B}}\simeq(0.1\sim 0.2)$ magnitude \cite{riess2011}. In Fig. \ref{f1}, we plot the 2-dimensional uncertainties contours for $M_{\rm B}$-$h_0$ and $M_{\rm B}$-$q_0$.  Note that the constraint on $q_0$ is weak by the low-$z$ observations, where we set  $q_0$ in the range of $q_0\in(-0.7,-0.4)$ in the analysis. In addition, we find that the constraints on the calibration parameters, as well as the cosmological parameters, are completely consistent with those of Cepheid calibration. For instance, we find that, although the uncertainty is still similarly large, the central value of $h_0$ becomes slightly larger than that derived from GW data alone \cite{GW170817Nature}, which supports the traditional Cepheid calibration method. This result hints that the tension between the SNIa and the CMB datasets in deriving the $h_0$ value still exists, which should not be caused by the Cepheid calibration method. If considering the SNIa data in a larger range, i.e. $z<0.2$, we find that the constraints on the parameters are also consistent in $1\sigma$ confidence level. However, in accordance to previous work \cite{SNIa-data}, we find that the central values of $b$ and $c$ become slightly smaller, which may give a hint on the evolution of the empirical relation of SNIa. We expect that future, and more precise, GW calibration can solve the issue.


\begin{table*}
\scriptsize
\caption{The constraints of calibrating parameters derived from various calibration analyses. Note that the super-index $a$ denotes the calibration method with Hubble law, and $b$ denote the calibration of interpolation analysis.}
\label{table1}
\begin{center}
\label{table2}
\begin{tabular}{|c| c| c| c| c| c| }
    \hline
     & ~~~~$a$~~~~ & ~~~~$b$~~~~ & ~~~~$c$~~~~ &~~~~$h_0$~~~~ &~~~~$q_0$~~~~\\
         \hline
    $^{a}$SNIa auto-calibration $z<0.05$ & $19.495^{+0.865}_{-1.010}$ & $0.148^{+0.010}_{-0.010}$ & $-2.836^{+0.180}_{-0.181}$ & $0.567^{+0.303}_{-0.297}$ &  $-0.565^{+0.104}_{-0.135}$  \\
         \hline
    $^{a}$GW170817/SNIa $z<0.05$ & $18.997^{+0.297}_{-0.322}$ & $0.148^{+0.008}_{-0.012}$ & $-2.836^{+0.121}_{-0.117}$ & $0.721^{+0.109}_{-0.096}$ &  $-0.566^{+0.103}_{-0.133}$  \\
    \hline
    $^{a}$GW170817/SNIa $z<0.1$ & $18.995^{+0.288}_{-0.304}$ & $0.150^{+0.008}_{-0.008}$ & $-2.919^{-0.109}_{-0.109}$ & $0.720^{+0.101}_{-0.095}$ &  $-0.593^{+0.091}_{-0.107}$  \\
    \hline
    $^{a}$GW170817/SNIa $z<0.2$ & $19.011^{+0.294}_{-0.320}$ & $0.139^{+0.006}_{-0.006}$ & $-3.058^{+0.077}_{-0.074}$ & $0.713^{+0.108}_{-0.097}$ &  $-0.545^{+0.082}_{-0.086}$  \\
    \hline
    $^{a}$3G network/SNIa $z<0.1$ & $19.122^{+0.009}_{-0.009}$ & $0.149^{+0.008}_{-0.008}$ & $-2.921^{+0.111}_{-0.112}$ & $0.673^{+0.002}_{-0.002}$ &  $-0.515^{+0.077}_{-0.082}$  \\
    \hline
    $^{b}$3G network/SNIa $z<0.1$ & $19.130^{+0.011}_{-0.011}$ & $0.150^{+0.010}_{-0.010}$ & $-2.784^{+0.139}_{-0.136}$ & ---------- &  ----------  \\ \hline
\end{tabular}
\end{center}
\end{table*}

\begin{table*}
\scriptsize
\caption{The constraints of calibrating parameters derived from various calibration analyses. Note that the super-index $a$ denotes the case with 1000 face-on GW events at $z>0.1$, and the super-index $b$ denotes the case with 5000 GW events with random inclination angles at $z>0.1$.}
\label{table2}
\begin{center}
\label{table2}
\begin{tabular}{|c| c| c| c| }
    \hline
     & ~~~~$a$~~~~ & ~~~~$b$~~~~ & ~~~~$c$~~~~ \\
    \hline
    $^{a}$3G network/SNIa $z>0.1$ & $19.107^{+0.005}_{-0.005}$ & $0.140^{+0.005}_{-0.006}$ & $-3.138^{+0.068}_{-0.068}$ \\
    \hline
    $^{a}$3G network/GRB $z>0.1$ & $52.735^{+0.014}_{-0.014}$ & $1.871^{+0.034}_{-0.035}$ & ----------   \\
    \hline
        $^{b}$3G network/SNIa $z>0.1$ & $19.111^{+0.007}_{-0.006}$ & $0.138^{+0.006}_{-0.006}$ & $-3.048^{+0.073}_{-0.072}$ \\
    \hline
    $^{b}$3G network/GRB $z>0.1$ & $52.748^{+0.016}_{-0.017}$ & $1.832^{+0.037}_{-0.038}$ & ----------   \\
    \hline
\end{tabular}
\end{center}
\end{table*}

\section{Calibrating cosmic ladders in the era of 3G GW detectors}



For the 3G GW detectors, the different kinds of noise will be reduced by more than one order of magnitude compared with the 2G ones, and the low-frequency range will be extended to sub-10 Hz \cite{et,ce}. Therefore, in addition to the much more precise observations on the low-$z$ BNSs, 3G network also provides the opportunity to detect the high-$z$ GW signals up to $z=2$ for BNSs and $z>2$ for NSBHs. Considering these sources, it then becomes possible to directly measure the absolute magnitude of SNIa, test the validity and/or calibrate the cosmic distance ladders in a wide range of redshifts for the first time. In this section, as an example, we forecast the calibration capabilities of the 3G network for both the SNIa and the GRB standard candles.

\subsection{Calibrating the SNIa standard candle in low-$z$ range}

Two leading proposals are currently under consideration for the design of 3G GW detectors. One is the ET \cite{et}, and the other is the CE \cite{ce}. We consider a 3G network consisting of both ET and CE. In order to calculate the uncertainties in the luminosity distance of GW sources, we adopt the Fisher matrix analysis \cite{zhao2017}. The response of an incoming GW signal is a linear combination of two wave polarizations in the transverse-traceless gauge, $d_I(t_0+\tau_I+t)=F_I^{+}h_+(t)+F_I^{\times}h_{\times}(t)$,
where $h_+$ and $h_{\times}$ are the plus and cross modes of GW respectively, $t_0$ is the arrival time of the wave at the coordinate origin, $\tau_I$ is the time required for the wave to travel from the origin and reach the $I$-th detector at time $t$. The detector's antenna beam-pattern functions $F_I^{+}$ and $F^{\times}_I$ depend on the source localization $(\theta_s,\phi_s)$, the polarization angle $\psi_s$, as well as the detector's location and orientation. For the GW signals $h_+$ and $h_{\times}$ of BNSs, we adopt the restricted post-Newtonian approximation of the waveform for the non-spining systems \cite{sathya}, including only waveforms in the inspiralling stage, which depends on the symmetric mass ratio $\eta=m_1m_2/(m_1+m_2)^2$, the chirp mass $\mathcal{M}_c=(m_1+m_2)\eta^{3/5}$, the luminosity distance $d$, the inclination angle of the binary system $\iota$, the merging time $t_c$ and merging phase $\psi_c$.

\begin{figure}
\begin{center}
\centerline{\includegraphics[width=9cm,height=7cm]{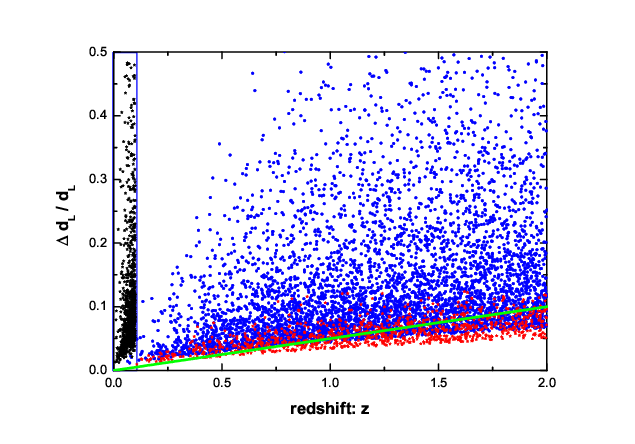}}
\end{center}\caption{Black dots denote the statistical distance uncertainties of 1000 low-$z$ GW events, red stars denote those of 1000 face-on high-$z$ GW events, and blue dots denote those of 5000 high-$z$ GW events with random inclination angles. The green line denotes the uncertainties caused by cosmic lensing. }\label{f2}
\end{figure}

For any given binary system, the response of the GW detector depends on nine system parameters ($\mathcal{M}_c,\eta,t_c,\psi_c,\theta_s,\phi_s,\psi_s,\iota,d$). Employing the 9-parameter Fisher matrix and marginalizing over the other parameters, we derive the uncertainty $\sigma_d$. In the low-$z$ range, the localization ability of 3G network is extremely high. For BNSs at $z=0.1$, the angular resolution is around $\Delta\Omega_s\in(1,10)$ deg$^2$ for 3G network \cite{zhao2017}. So, it seems possible to identify the EM counterparts for a large number of GW events. We numerically simulate the BNS samples
with random binary orientations and sky directions for our investigations, and assume that the mass of each NS is $1.4~{\rm M}_{\odot}$. The redshifts are uniformly distributed in comoving volume in the range $z<0.1$, considering a spatial flat $\Lambda$CDM cosmology with $h_0=0.7$ and $\Omega_{m}=0.3$. For each sample, we calculate the values of SNR and $\Delta d/d$. Note that the distance uncertainty, $\sigma_d/d$, of GW sources are subject to two kinds: the statistical error $\Delta d/d$, and the additional error due to the effects of weak lensing which should also be considered, and it can be approximated as $\tilde{\Delta} d/d=0.05z$ \cite{sathya2009}. We randomly select 1000 sources to mimic the detection of 3G network in three years \cite{GW170817PRL}, which satisfy the criteria of ${\rm SNR}>8$, and $\Delta d/d<50\%$. The distribution of $\Delta d/d$ for these samples are presented in Fig.\ref{f2}. Using these samples, we calibrate the empirical relation of SNIa in the low-$z$ with $z<0.1$ by two different methods. The first one is the same as described previously, where we use both SNIa data and the simulated GW data to simultaneously constrain five parameters ($a$,$b$,$c$,$h_0$,$q_0$). In the second method, we consider the linear interpolation of GW sources to obtain the value of $S_i$ at each SNIa redshift and the corresponding error bar. We apply the following $\chi^2$ calculation to obtain the constraints on $a$, $b$, $c$,
\begin{equation}\label{eq1}
{\chi}_{\rm SN}^2=\sum_{i=1}^{N}\frac{(S_i-a-bX_i-cY_i-Z_i)^2}{\sigma_{S_i}^2+\sigma^2_{a+bX_i+cY_i+Z_i}}.
\end{equation}
We present the calibration results in Table \ref{table1}, where we show that they are nearly the same for both methods. The uncertainties of the absolute luminosity $M_{\rm B}$ are $30$ times smaller than the present ones, which is also nearly one order of magnitude smaller than the result of Cepheid calibration. Note that although the central value of the calibrated parameters depends on the assumed fiducial cosmological model, their uncertainties are nearly independent of it,  thus quantifying the calibration capabilities of the corresponding methods.

\subsection{Determining the absolute magnitude of SNIa in high-$z$ range}

Observing the high-$z$ GW events by 3G detector network provides a unique way to calibrate the cosmic ladders in the high-$z$ range, which is crucial for SNIa. Since in the standard way, the empirical relation is calibrated in the low-$z$ range by Cepheid variables, and directly extrapolated to more than $500$ farther distances range. The principle problem of this extension is their ``standardness" of the brightest standard candles: How homogeneous the objects are in their true absolute magnitude. For some of these different standard candles, the homogeneity is based on theories about formation and evolution of stars and galaxies, and it is thus also subject to uncertainties in those aspects. For the SNIa, the most luminous of the distance indicators, this homogeneity is known to be poor \cite{poor}. However, no other class of object is bright enough to be detected at such large distances, so the class is useful simply because there is no real alternative. Fortunately, the GW detection of BNS and NSBH combined with their EM counterparts, could provide an independent method to test and calibrate these candles. In addition to correct the possible biases caused by the high-$z$ SNIa, the new calibration is also helpful to study the intrinsic physics of SNIa models, for instance, the effect of metallicity and local strength of gravity on the bursts.

It has been argued that for high-$z$ binary coalescent GW events, the most promising method to measure their redshifts is to observe their short-hard $\gamma$-ray burst (shGRB) counterparts and the afterglows. However, the angular distributions of these EM signals are still unclear \cite{unclear,unclear2,shGRB}. In this paper, we consider two extreme cases. Similar to the previous works \cite{sathya2009,zhao2011,zhao2017}, in the first case, we assume the shGRBs are beamed: The $\gamma$ radiation is emitted in a narrow cone more or less perpendicular to the binary orbital plane, and the observed shGRBs are nearly all beamed towards the Earth \cite{shGRB}. In the second extreme case, we assume the energetics of shGRBs is isotropic. Therefore, in the first case, only the nearly face-on GW events can be detected by both GW and EM bands, and the expected event rate in this case is low. For a conservative estimation, as in previous works \cite{zhao2017,sathya2009,zhao2011}, we assume 1000 observed BNSs uniformly distributed in comoving volume for a redshift range of $0.1 < z < 2$. In calculation, we randomly choose 1000 samples from the full simulations, which satisfy the criteria of ${\rm SNR}>8$. These samples mimic the potential observation of 3G network, and are used to calibrate the cosmic ladders. While in the second case, the inclination angles of observed GW events with EM counterparts are expected to be randomly distributed, and the event rate in this case is expected to be much higher. In this paper, for a conservative estimation, we assume 5000 observed BNSs satisfying the criteria of ${\rm SNR}>8$, and $\sigma_d/d<50\%$, which are randomly chosen from the full simulations. For these high-$z$ samples, the source localizations $(\theta_s,\phi_s)$ of GW events are expected to be fixed by their EM counterparts, which can be excluded from the Fisher matrix analysis. Utilizing these samples, we repeat the same calculation as stated above, but adopting now a 7-parameter Fisher matrix, and find that the statistical uncertainties $\Delta d/d$ is quite large. For instance, in the second case, if $z>0.5$, more than $60\%$ samples have $\Delta d/d>50\%$. In the first case, the situation is even worse: nearly all the samples have $\Delta d/d>1$. These results are consistent with that in our previous work \cite{zhao2017} (see. Fig. 14), which are caused by the strong degeneracy between the luminosity distance $d$ and the inclination angle $\iota$ in GW analysis. So, if the inclination angle of GW event is unknown, the application of BNSs as standard sirens in high-$z$ range is limited.

For this reason, similar to the recent work \cite{sathya2019}, in this article, we consider the scenario of assuming that the inclination angles, as well as the sky-positions, of BNSs are known purely from their EM counterparts. This scenario is possible as we have already have seen in the case of GW170817. The sky position of GW170817 was constrained by finding the host galaxy NGC 4993 \cite{host} whereas the inclination angle was constrained from the X-ray and ultraviolet observations \cite{fix-iota}. In this scenario, the errors in distance measurement can be significantly reduced. For both cases, we calculate the value of $\Delta d/d$ by analyzing the 6-parameter Fisher matrix for each sample, and present their distribution in Fig. \ref{f2}. We find that the values of $\Delta d/d$ in case two is larger than those of case one by a factor 1-2, which is caused by the higher SNR for the face-on GW events.

We apply these samples to calibrate the empirical relation of SNIa $S=a+bX_1+c\mathcal{C}+Z$ as above, and assume the homogeneity of this empirical relation in the full range $0.1<z<2$. In the high-$z$ range, the Hubble law is untenable and cannot be applied anymore. Therefore, for different redshift ranges, we use a linear interpolation of GW sources to obtain the value of $S_i$ at the SNIa redshift and the corresponding error bar. We apply the $\chi^2$ calculation defined in (\ref{eq1}) to obtain the constraints on $a$, $b$ and $c$, which are listed in Table \ref{table2} for both cases. We find the uncertainties of all parameters $a$, $b$ and $c$ in case two are slightly larger than those in case one, which is caused by the larger uncertainties $\Delta d/d$ in case two.

Note that the numbers of observable GW events, $N_{\rm GW}$, in both cases are quite uncertain. In order to test how the uncertainties of parameters $a$, $b$ and $c$ depend on $N_{\rm GW}$, in case one, we change the value of $N_{\rm GW}=1000$ to that of $N_{\rm GW}=300$, and repeat the above calculation. We find the results in two cases are nearly the same, although the values of $N_{\rm GW}$ are quite different. This is is understandable: The GW observations influence our results only through affecting the values of $S_i$ and their uncertainties of SNIa samples. Since we consider the linear interpolation of GW sources to obtain the value of $S_i$ at each SNIa redshift and the corresponding error bar, and their values depend only on its nearby GW events. Therefore, increasing or decreasing $N_{\rm GW}$ cannot significantly affect their values. In case two with GW samples with random inclination angles, we find the similar results: Changing $N_{\rm GW}$ from 5000 to 2000, or to 20000, we cannot find the significant changes on the uncertainties of  parameters $a$, $b$ and $c$.

From Table \ref{table2}, we find that the constraints on $b$ and $c$ are similar to the current constraint level, which is dominated by the observed SNIa data. However, the uncertainty of $M_{\rm B}$ is around $0.006$ magnitude, which is more than one order tighter than the Cepheid calibration in the low-$z$ range. These results demonstrate that the potential GW standard sirens indeed provide a powerful tool to calibrate the empirical relation of SNIa in the full redshift range.

\subsection{Calibrating the empirical relation of GRBs}

In cosmic distance ladder, the GRBs have been proposed as a complementary probe, which provide a new tool to measure the distance of objects in higher redshifts up to $z\sim 8$, due to their intense explosions \cite{grb}. An important concern is related to the calibration of the empirical relation between luminosity and energy of the GRBs, which is always carried on by assuming the SNIa as standard candles and using the observed SNIa data for $z<1.4$ \cite{lang2008}. However, the GW observations on the coalescing BNSs and NSBHs provide another model-independent method to calibrate these relations. Here, we investigate the potentials of GW calibration on the empirical relation of GRBs.

For GRB data, we adopt the most recent observation, including 81 GRB data at $z_{\max}<2.0$ summarized in \cite{WeiHao2015}. For each GRB data, the empirical Amati relation reads $S=a+bX+Y+Z$, where $S=2\log_{10}(d/{\rm cm})$, $X=\log_{10}\left({E_{p,i}}/{300{\rm keV}}\right)$ with $E_{p,i}$ being the cosmological rest-frame spectrum peak energy of GRBs, $Y=-\log_{10}(S_{\rm bolo}/{10^{-5}~{\rm erg}~{\rm cm}^2})$ with $S_{\rm bolo}$ being the bolometric fluence of gamma rays in the GRB, and $Z=5-\log_{10}(4\pi/(1+z))$. Similar to the discussion above, the values of $S_i$ are calculated from the linear interpolation of GW sources, and the $X_i$, $Y_i$ and $Z_i$ are the observed data. Note that the calibration parameters $a$ and $b$ correspond to the parameters $\lambda$ and $b$ in the literature \cite{grb,lang2008,WeiHao2015}. By a similar $\chi^2$ analysis as in the previous discussion, we obtain the constraints on the calibration parameters $a$ and $b$, which are listed in Table \ref{table2}. We find that their uncertainties are one order of magnitude smaller than the current constraints derived from the SNIa calibration \cite{lang2008}.

~

\section{Conclusions}

The detection of GW170817 by the collaboration of two Advanced LIGO detectors and one Advanced Virgo detector opens the possibility of using a brand new GW window to further our understanding of fundamental physics, cosmology and astrophysics. Due to the self-calibrated distance of GW sources, GW can be treated as a standard siren to independently construct the Hubble diagram in a wide range of redshifts, by which the absolute magnitude $M_{\rm B}$ of SNIe at different redshifts can be directly measured. Considering the 2G GW detector network, the measurement of $M_{\rm B}$ at low-$z$ range provides a novel way to calibrate the empirical Phillips relationship of SNIa. In this paper, we calibrate this relation using the observation of GW170817. We find that the calibration result is consistent with that derived from traditional Cepheid calibration, and the uncertainties of $M_{\rm B}$ are comparable in both methods. For the calibration capability of 3G GW detector network, we find that the uncertainty of $M_{\rm B}$ in both low-$z$ and high-$z$ is more than one order of magnitude smaller than that from the present level, which provides a unique opportunity for research on cosmology, gravity and intrinsic properties of SNIa at high redshifts. In addition, the empirical Amati relation of high-$z$ GRBs can also be calibrated, and the uncertainties of the parameters are expected to be one order of magnitude smaller than those derived from traditional SNIa calibration.

\begin{acknowledgments}
We appreciate the helpful discussions with Linqing Wen. This work is supported by NSFC Grants Nos. 11773028, 11603020, 11633001, 11173021, 11322324, 11653002 and 11421303, the project of Knowledge Innovation Program of Chinese Academy of Science, the Fundamental Research Funds for the Central Universities and the Strategic Priority Research Program of the Chinese Academy of Sciences Grant No. XDB23010200.
\end{acknowledgments}


\baselineskip=12truept

\begin{thebibliography}{99}

\bibitem{SNIa-review}
W. Hillebrandt and J. C. Niemeyer, ARA\&A, {\bf 38}, 191 (2000).

\bibitem{riess2009}
A. G. Riess, et al., ApJS, {\bf 183}, 109 (2009).

\bibitem{phillipse}
M. M. Phillipse, ApJ, {\bf 413}, L105 (1993).

\bibitem{ceipheid-review}
W. L. Freedman and B. F. Madore, ARA\&A, {\bf 48}, 673 (2010).

\bibitem{poor}
M. Gilfanov and A. Bogdan, Nature, {\bf 463}, 924 (2010).

\bibitem{double}
B. E. Shaefer and A. Pagnotta, Nature, 481, 164 (2012).

\bibitem{LiBaojiu2017}
B. S. Wright and B. Li, Phys.~Rev.~D {\bf 97}, 083505 (2018).



\bibitem{zhao2018}
W. Zhao, B. S. Wright and B. Li, JCAP {\bf 10}, 052 (2018).



\bibitem{schutz}
B. Schutz, Nature, {\bf 323}, 310 (1986).

\bibitem{shGRB}
E. Nakar, Phys. Rep., {\bf 442}, 166 (2007).


\bibitem{et}
http://www.et.et-gw.eu/.

\bibitem{ce}
B. P. Abbott et al., Class. Quantum Grav., {\bf 27}, 084007 (2010).

\bibitem{sathya2019}
A. Gupta, D. Fox, B. S. Sathyaprakash and B. F. Schutz, arXiv:1907.09897.

\bibitem{wangfy}
Y. Y. Wang and F. Y. Wang, ApJ {\bf 873}, 39 (2019).

\bibitem{planck2018}
Planck Collaboration, arXiv:1807.06209.

\bibitem{riess2019}
A. G. Riess,  S. Casertano, et al., arXiv:1903.07603.


\bibitem{GW170817PRL}
LIGO Scientific Collaboration and Virgo Collaboration, Phys. Rev. Lett., {\bf 119}, 161101 (2017).

\bibitem{GW170817Nature}
B. P. Abbott, et al., Nature {\bf 551}, 85 (2017).

\bibitem{SNIa-data}
M. Betoule, et al., A\&A, {\bf 568}, A22 (2014).

\bibitem{weinberg}
S. Weinberg, Cosmology, (Oxford University Press, New York, 2008).


\bibitem{riess2011}
A. G. Riess et al., ApJ {\bf 730}, 119 (2011).

\bibitem{zhao2017}
W. Zhao and L. Wen, Phys. Rev. D {\bf 97}, 064031 (2018).

\bibitem{sathya}
B. S. Sathyaprakash and B. Schutz, Living Reviews in Relativity, {\bf 12}, 2 (2009).

\bibitem{sathya2009}
B. S. Sathyaprakash, B. Schutz and C. van den Broeck, Class. Quantum Gravity, {\bf 27}, 215006 (2010).


\bibitem{unclear}
LIGO Scientific Collaboration, Virgo Collaboration, Fermi Gamma-Ray Burst Monitor, INTEGRAL, ApJ {\bf 848}, L13 (2017).

\bibitem{unclear2}
M. Kasliwal, E. Nakar, L. P. Singer et al., Science {\bf 358}, 1559 (2017).

\bibitem{host}
LIGO Scientific Collaboration and Virgo Collaboration, ApJ {\bf 848}, L12 (2017).

\bibitem{fix-iota}
P. A. Evans, et al., Science {\bf 358}, 1565 (2017).


\bibitem{zhao2011}
W. Zhao, C. van den Broeck, D. Baskaran and T. G. F. Li, Phys. Rev. D, {\bf 83}, 023005 (2011).


\bibitem{grb}
B. E. Schaefer, ApJ, {\bf 660}, 16 (2007).

\bibitem{lang2008}
N. Lang, W. K. Xiao, Y. Liu and S. N. Zhang, ApJ, {\bf 685}, 354 (2008).

\bibitem{WeiHao2015}
J. Liu and H. Wei, Gen. Rel. Grav. {\bf 47}, 141 (2015).


\end{thebibliography}
\end{document}